\documentclass[12pt,oneside,reqno]{amsart}
\usepackage{geometry} 
\usepackage{amsfonts,amsmath,amssymb,amsthm,amscd,
latexsym,graphicx,subfigure,xcolor}
\renewcommand{\vec}[1]{\boldsymbol{#1}}
\usepackage{amsaddr}
\usepackage{bm}

\begin{document}

\title{Qualitative considerations in Intense Field QED}
\author{A.M. Fedotov}
\address{National Research Nuclear University ``MEPhI'' \\(Moscow Engineering Physics Institute)}

\begin{abstract}
I demonstrate how at least some of the known results on probability rates of the basic processes in strong external field could be understood in a simple-man fashion accessible to a high-school student, by using merely only kinematical considerations combined with the uncertainty principle. 
\end{abstract}

\maketitle

\section{Introduction}

\begin{figure}[th!]
\includegraphics[width=\columnwidth]{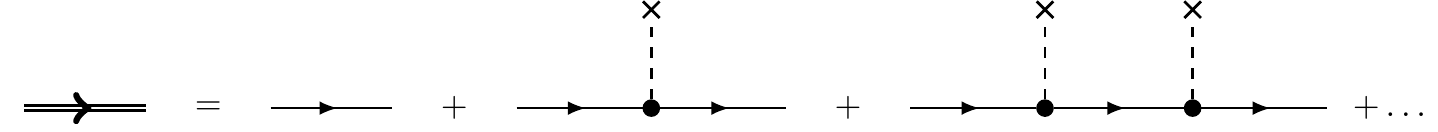}
\caption{\label{Fig1} `Dressed' electron propagator in external field.}
\end{figure}

Intense Field QED (IFQED) is nowadays, at least theoretically, a well-developed research area with more than 60 years of history, see e.g. recent reviews \cite{diPiazza12,Narozhny15} and references therein. In practice, IFQED implies the almost traditional QED calculations by using the ordinary rules of the Feynman diagram technique, but with all the free electron propagators or external lines replaced with the exact (sometimes also called `dressed') ones, which take into account interaction with an external field to all orders exactly (see Fig.~\ref{Fig1}). An exact (or `dressed') propagator or external line can be constructed explicitly once the Dirac equation in a corresponding external field is exactly solvable. 

IFQED is interesting on its own rights as a non-perturbative QFT model, but has also wide range of applications to astrophysics. Interest to IFQED effects essentially increased after the SLAC E144 experiment \cite{SLAC1,SLAC2}, and especially in a view of expectations for further growth of the available laser intensity aimed by such recent state-of-the-art projects as Extreme Light Infrastracture (ELI) or EXawatt Center for Extreme Light Studies (EXCELS).

However, in most of the original papers or textbooks on the subject the results are usually obtained by {\it extremely bulky} analytical, or since recently by even numerical, calculations. On the one hand, merely everybody would agree that {\it qualitative considerations} always allow to gain deeper insights into a problem. But surprisingly, on the other hand qualitative considerations in IFQED have been almost never discussed in the literature in general setting. Among just a few notable exceptions a couple of papers \cite{Migdal72,Akhmedov11} could be mentioned. However, each of them covers only a selected aspect of IFQED. In my opinion, elaboration of physical intuition in such kind of problems is now becoming a real challenge and would be of great importance for further developments of the theory.

In this paper, I am trying to reproduce the rather well known asymptotic expressions for probability rates of the simplest IFQED processes involving ultrarelativistic particles (hard photon emission, pair creation, mass operator) without even a bit of explicit calculations. The tools in use are essentially only the kinematical arguments and the uncertainty principle, so that these results can be now explained for the first time to a high-school student. Deeper understanding of the role and interplay of characteristic length/time scales of the problem also comes as a byproduct of the approach.

\section{The approach}

Let us start with more or less general settings. Consider a process of particle interaction in a strong external field. During the process, some initial ingoing particles start interacting with each other, as well as with an external field, and finally transform into the (same or different) final collection of definite outgoing particles. In QED such a process is described by one or more Feynman diagram. Assume for a moment, that the external field is switched on and off adiabatically. In the Dyson-Feynman formalism, all the initial particles are on-shell, while in the absence of the field energy and momentum are conserved at all the interaction vertices, and the process takes place if all the outgoing particles also go on-shell. 

However, other equivalent description is also possible (and historically preceded, see e.g. Ref.~\cite{Schwartz14}), for which all the particles, including virtual, are on-shell, momentum is conserved at the vertices, but instead energy is not conserved. This is the so-called `old-fashion' perturbation theory, the same one that is well known from the non-relativistic quantum mechanics. Such a description is now rather rarely used in relativistic problems, because it treats energy and momentum differently and thus is not manifestly relativistic covariant (though, yet again, it is equivalent and hence covariant implicitly). In addition, it is also ineffective practically, since the now usual Dyson-Feynman approach automatically combines together different terms of old-fashion PT. However, as long as we are interested just in introducing some useful definitions rather than in literal practical calculations, we prefer to use it in discussion here.

If the energy of outgoing particles exceeds the one of ingoing particles (energy lack $\Delta\varepsilon=\sum \varepsilon_f-\sum\varepsilon_i>0$), then the process can not take place in the absence of external field. However, in its presence particles can gain the lacking energy from the field and (provided it is not forbidden by other remaining conservation laws) the process becomes possible. Such sort of processes are often called laser-induced. 

Let us now define two characteristic times associated with such a process. The first one is related to $\Delta\varepsilon$ via the uncertainty principle\footnote{We use units $\hbar=c=1$ throughout the paper.}, 
\begin{equation}\label{tq}
t_q\simeq\frac{1}{\Delta\varepsilon}.
\end{equation}
As known, this principle states that energy lack $\Delta\varepsilon$ should be redeemed within the time $t\lesssim t_q$. The second time $t_e$ is the one that it takes for particles to gain energy $\simeq \Delta\varepsilon$ from the field,
\begin{equation}\label{te}
e\int\limits_0^{t_e}\vec{E}\cdot d\vec{s}\simeq\Delta\varepsilon.
\end{equation}

As just stated above, the field-induced processes are classically forbidden.
Obviously, if $t_e\lesssim t_q$, then the process is allowed, let us call such case the `quantum regime'. Otherwise ($t_e\gtrsim t_q$) the process should be somehow suppressed, let us call it `quasiclassical regime'.

\begin{figure}[th!]
\includegraphics[width=0.5\columnwidth]{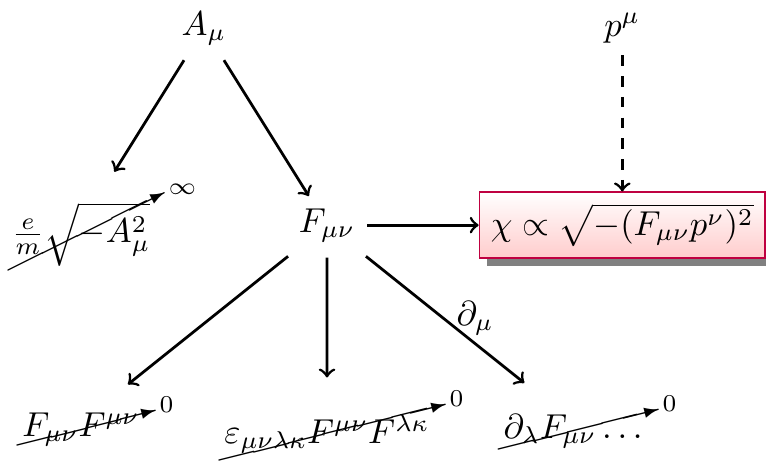}
\caption{\label{Fig2} Diagram illustrating the constant crossed field approximation: the exclusive probability of the process can depend only on the field and the $4$-momentum of initial particle. Under our conditions there remains just a single Lorentz- and gauge-invariant parameter $\chi$.}
\end{figure}

To quantify the suppression factor in quasiclassical regime let us consider a special case of constant purely electric field, described by vector potential $\vec{A}(t)=-\vec{E}t$. We assume throughout the paper that $E\ll E_S=m^2/e=1.3\times 10^{16}$V/cm. Note that such an example is of course special. However, if the field frequency is small (see below), there is just a single initial particle and all the participating particles are ultrarelativistic, having non-small angle between the direction of propagation and the Poynting vector of the field, then any field looks as constant and crossed ($\vec{E}\cdot\vec{B}\approx 0$, $E\approx B$) and hence can be equivalently replaced with a constant electric field of the same value of the dimensionless quantum parameter 
\begin{equation}\label{chi}
\chi=\frac{e}{m^3}\sqrt{-(F_{\mu\nu}p^\nu)^2}=\frac{\gamma\sqrt{\left(\vec{E}+\vec{v}\times\vec{H}\right)^2-(\vec{v}\cdot\vec{E})^2}}{E_S}=\frac{E_P}{E_S},
\end{equation}
where $E_P$ is the electric field strength in the proper reference frame. For electron, the parameter $\chi$ is its proper acceleration in Compton units.
The essence of this important approximation is explained in more details in Ref.~\cite{Ritus64} and illustrated in Fig.~\ref{Fig2}.
In fact, we will mostly discuss below just this latter case, because it is of most importance for laser-matter interactions at ultrahigh intensities.

If $E\ll E_S$ (no spontaneous pair creation from vacuum), then the solution of Dirac equation in the gauge specified above can be {\it approximately} cast to the form
\begin{equation}\label{qc_sol}
\Psi(\vec{r},t)\propto \exp\left\{i\vec{p}\vec{r}-i\int\limits_0^t \varepsilon_{\vec{p}}(t')\,dt'\right\},\; \varepsilon_{\vec{p}}(t)=\sqrt{\left(\vec{p}-e\vec{A}(t)\right)^2+m^2}.
\end{equation}
It is of the quasiclassical form $e^{iS}$, as well as is the (but in that case exact) Volkov solution in a constant crossed field. Obviously, the quantum transition amplitude with such solutions would always take the form
\begin{equation}\label{tr_ampl}
c_{i\to f}=-i\int\limits_{-\infty}^{+\infty}dt\,V_{fi}(t)\exp\left\{i\int\limits_0^t \Delta\varepsilon(t')\,dt'\right\},\quad V_{fi}(t)\propto\delta^{(3)}\left(\Delta\vec{p}\right).
\end{equation}
Exact conservation of generalized momentum is in fact our motivation to prefer the constant electric field. In a regime when the transition is suppressed, this integral can be estimated by the saddle-point method:
\begin{equation}\label{tr_amp_est}
c_{i\to f}\propto \exp\left\{-\int\limits_0^{t_*} \Delta\varepsilon(it')\,dt'\right\},
\end{equation}
where $t_*$ is the smallest positive solution to the equation $\Delta\varepsilon(it_*)=0$. Together with $\Delta\vec{p}=0$ (see Eq.~(\ref{tr_ampl})) this condition defines the endpoint of quantum tunneling\footnote{In the picture adopted here, this is tunneling in time rather than in space. The same idea underlies the imaginary time method  \cite{Popov} and, in a more refined way, the instanton technique.}. As we will observe below on particular examples, as a rule $t_*\simeq t_e$. Hence Eq.~(\ref{tr_amp_est}) roughly means that $c_{i\to f}=\mathcal{O}(e^{-t_e/t_q})$.

Let us now illustrate the approach with the simplest (and perhaps the only one already well understood qualitatively) example of spontaneous pair production in a constant electric field. For the sake of simplicity, let us assume from the beginning that the pair is created at rest ($\vec{p}_\perp=0$ for both particle and antiparticle -- this matches automatically momentum conservation). In such a case $\Delta\varepsilon=2m$ and $t_e\simeq \Delta\varepsilon/eE\simeq m/eE$, while $t_q\simeq 1/\Delta\varepsilon\simeq 1/m$. Since for $E\ll E_S$ we always have $t_e\gg t_q$, we expect the process to be suppressed exponentially as $\mathcal{O}(e^{-t_e/t_q})=\mathcal{O}(e^{-E_S/E})$. In order to state this more precisely, we take into account that we have vacuum at the beginning and a pair in the final state to write
\[\Delta\varepsilon(t)=2\varepsilon_{\vec{p}=0}(t)=2\sqrt{m^2+e^2E^2t^2}\]
and solve $\Delta\varepsilon(it_*)=0$ for $t_*$ to get $t_*=m/eE$. Note that in this case $t_*=t_e$, as announced above. Now we can compute
\begin{equation}\label{Schwinger}
W_{e^-e^+}=\left\vert\exp\left\{-2\int\limits_0^{m/eE}\sqrt{m^2-e^2E^2{t'}^2}\,dt'\right\}\right\vert^2=e^{-\pi m^2/eE},
\end{equation}
which coincides exactly with the known result.

Let us note that the `quantum time' and more generally `the formation time/length' is obviously defined non-uniquely. As a result, there is discrepancy in the literature in definition of the proper `formation time/length' even for this simplest process. For example, instead of energy lack we could derive it via uncertainty relation from the work produced by the field, $eE\tilde{t}\simeq 1/\tilde{t}$, $\tilde{t}\simeq 1/\sqrt{eE}$. For $E\ll E_S$ we have $t_e\gg \tilde{t}_q\gg t_q$. This time seems to be also important, for example the pre-exponential factor can be cast to a form $e^2E^2VT/(4\pi^2)\simeq VT/\tilde{t}^4$, i.e. interpreted as exactly the number of loops in the space-time region occupied by the field, if each loop is thought to be exactly of size $\tilde{t}_q$ (see also \cite{Migdal72}). On the other hand, Nikishov \cite{Nikishov79} defines the `formation time' of pair creation as a time interval of non-validity of the quasiclassical solution (\ref{qc_sol}), $t_N\simeq m^{-1} (E_S/E)^{3/2}$. For $E\ll E_S$, this is the largest time among all the possibilities considered above. Our approach is not to focus on some single time, but rather to consider and compare two characteristic times $t_e$ and $t_q$, both of which obviously have a clear physical meaning. Now let us try to apply it to some more complicated processes, those not have been considered yet qualitatively at all.

\section{Hard photon emission}
\label{Sec3}

\begin{figure}[th!]
\centering
\subfigure[]{
\includegraphics[width=0.2\textwidth]{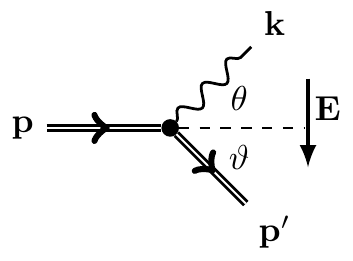}
\label{hpe-feynman}}
\subfigure[]{
\includegraphics[width=0.4\columnwidth]{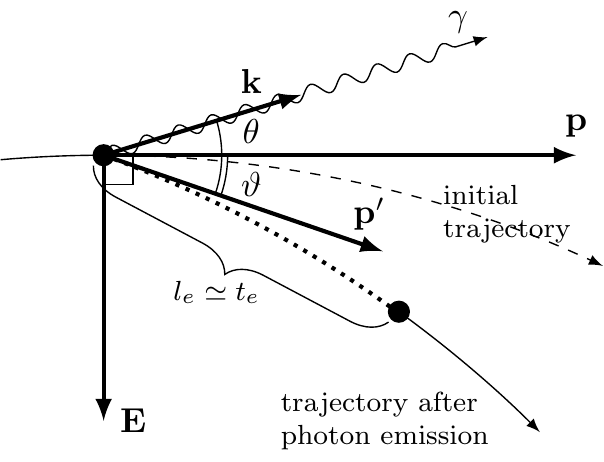}
\label{hpe-pic}}
\caption{\label{Fig3} Hard photon emission in a constant electric field: Feynman diagram \subref{hpe-feynman}, and a more detailed sketch of quasiclassical trajectories \subref{hpe-pic}. The dashed path is the continuation of an initial trajectory and the dotted path represents virtual electron after the recoil.}
\end{figure}

Let us start with the process of hard photon emission, also often called the non-linear Compton scattering or the classical/quantum synchrotron radiation (see Fig.~\ref{Fig3}). In what follows, we assume that the field carrier frequency is low enough to consider the field constant (see below for a precise restriction). We also assume that electron is ultrarelativistic before ($p\gg m$) and after ($p-k\gg m$) photon emission and moves not at small angle with respect to the direction of the Poynting vector of the field. In such a case, an arbitrarily shaped field can be replaced by a constant crossed field, provided that the parameter $\chi$ keeps the same value. For simplicity, we replace it further by a constant electric field with $\vec{A}=-\vec{E}t$, directed perpendicularly to the initial momentum of the electron (still keeping the value of $\chi$ the same). 

The initial energy of the electron is then given by 
\begin{equation}\label{in_e}
\varepsilon_{\vec{p}}(t)=\sqrt{(\vec{p}-e\vec{A})^2+m^2}=\sqrt{p^2+e^2E^2t^2+m^2}
\approx p+\frac{e^2E^2t^2+m^2}{2p},
\end{equation}
where we have made an expansion assuming $p\gg m$ and $p\gg eEt$ (the latter to be justified by the result). 

For the outgoing electron, we would have
\begin{equation}\label{pf}
\vec{p}'=\vec{p}-\vec{k},\quad 
{p'}^2=p^2+k^2-2pk\cos{\theta}=(p-k)^2+4pk\sin^2(\theta/2).
\end{equation}
Let us assume (again, just for the sake of simplicity) that the process is planar (i.e., all the vectors $\vec{p},\vec{k}$ and $\vec{E}$ lie in the same plane). Then the final energy of the electron is given by
\begin{eqnarray}
\varepsilon_{\vec{p}'}(t)=\sqrt{(\vec{p'}-e\vec{A})^2+m^2}=\sqrt{{p'}^2-2e\vec{E}\cdot\vec{k}t+e^2E^2t^2+m^2}\approx\nonumber\\
\approx p-k+\frac{e^2E^2t^2+2eEkt\sin{\theta}+m^2+4pk\sin^2(\theta/2)}{2(p-k)}\label{ef}
\end{eqnarray}

Using (\ref{in_e}) and (\ref{ef}), we can represent the total energy lack for this process in the form
\begin{equation}\label{de_hpe}
\Delta\varepsilon(t)=\varepsilon_{\vec{p}'}(t)+k-\varepsilon_{\vec{p}}(t)
\approx\frac{k\left[e^2E^2t^2+2eEpt\sin{\theta}+m^2+4p^2\sin^2(\theta/2)\right]}{2p(p-k)}.
\end{equation}
Note that the leading terms have been perfectly cancelled, so that even if $k\sim p$ we have $\Delta\varepsilon=\mathcal{O}(m/\gamma)$, i.e. the outgoing virtual electron is almost real from the beginning. This seems to be the general property of the processes involving ultrarelativistic particles. Note also that up to now all our calculations were just purely kinematical.

Now to proceed further let us consider the limiting cases. Let us first assume that $t\lesssim m/eE$ (the weak field limit). Then we can skip the first two terms in the numerator of Eq.~(\ref{de_hpe}) against the remaining terms. Obviously, since the process is laser induced and the field is assumed to be weak, the main contribution to the process should come with the smallest possible energy lack. If $k$ is fixed, this means that the last term in the numerator of Eq.~(\ref{de_hpe}) should not exceed essentially the remaining term. This means that $\theta\lesssim m/p=1/\gamma$, i.e. we recover the standard property of decays of ultrarelativistic particles to emit inside a narrow cone of aperture $\mathcal{O}(1/\gamma)$ around the propagation direction. Since we are interested just in an estimation, we can now drop this last term as well by replacing strict equality by the approximate one,
\begin{equation}\label{de_hpe1}
\Delta\varepsilon\approx\frac{m^2}{2p(p-k)}.
\end{equation}
Then for the case under consideration we have
\begin{equation}\label{tq_hpe}
t_q\simeq \frac1{\Delta\varepsilon}\approx\frac{2p(p-k)}{m^2}.
\end{equation}

Estimation of $t_e$ is a bit more delicate because of the following reason. Due to momentum conservation, $k_\perp=p_\perp'$, and because also $p'\approx p-k$, we have $\vartheta=k\theta/p'\simeq km/[(p-k)p]\ll 1$. This is in full agreement with relativistic kinematics but means that production of work by the field is supressed, because virtual electron moves almost perpendicularly to the field. Taking this into account, we arrive at

\begin{equation}\label{te_hpe}
t_e\simeq\frac{\Delta\varepsilon}{eE\vartheta}\simeq\frac{m}{eE}.
\end{equation}
The process would be allowed if $t_q\gtrsim t_e$, otherwise we expect it would be suppressed exponentially. This matches precisely the known position of the exponential cutting edge of the classical synchrotron radiation spectrum, 
\begin{equation}\label{k_hpe}
k\lesssim k_{max}=\frac{eEp}{m}p=\chi p.
\end{equation}
Since $k\lesssim p$, this means that we are in the classical regime $\chi\lesssim 1$.
The probability rate $W_\gamma$ of the hard photon emission is proportional to $e^2$ (since there is just a single vertex) and has dimension $[\text{time}]^{-1}$. Hence, using the smallest time scale of the proplem ($t_e$), we can estimate it by
\begin{equation}\label{Wp}
W_\gamma\simeq \frac{e^2}{t_e}\sim \frac{e^2m^2}{p}\chi.
\end{equation}
By using (\ref{k_hpe}), (\ref{Wp}) we can now estimate also the average radiation reaction force acting on the electron,
\begin{equation}\label{Frr}
F_{RR}\simeq kW_\gamma\sim e^2m^2\chi^2.
\end{equation}
Eqs.~(\ref{Wp}), (\ref{Frr}) should be compared with the exact results \cite{Ritus64,Baier67,Landau4}
\begin{equation}\label{hpe_exact}
W_\gamma\approx \frac{5}{2\sqrt{3}}\frac{e^2m^2}{p}\chi\approx 1.443\frac{e^2m^2}{p}\chi,\quad F_{RR}\approx \frac23e^2m^2\chi^2.
\end{equation}

Now let us discuss briefly the much more interesting strong field limit $t\gg m/eE$. Now the first term in the numerator of Eq.~(\ref{de_hpe}) becomes dominant, $k\sim p$ and the allowed emission angles (those that do not increase the energy lack $\Delta\varepsilon$ too much) are given by 
\begin{equation}\label{angles}
\vartheta,\theta\lesssim\frac{eEt}{p}\ll 1.
\end{equation}
Obviously, now they originate mainly due to contortion of the trajectory of a virtual electron by the field. As for the energy lack, this time it is given by
\begin{equation}\label{de_hpe2}
\Delta\varepsilon(t)\simeq \frac{e^2E^2t^2}{p}.
\end{equation}
Interestingly, because of (\ref{angles}), the relation $eE\vartheta(t)t\simeq \Delta\varepsilon(t)$ is now satisfied identically, which must mean that for $t\gtrsim m/eE$ the particle has always enough time to gain the lacking energy from the field to become real.

The `quantum time' $t_q$ is defined by
\begin{equation}\label{tq_hpe1}
t_q\simeq\frac1{\Delta\varepsilon(t_q)}\simeq\frac{p}{e^2E^2t_q^2},
\end{equation}
hence we have
\begin{equation}\label{tq_hpe2}
t_q\simeq\left(\frac{p}{e^2E^2}\right)^{1/3}=\frac{p}{m^2}\chi^{-2/3}=\frac{m}{eE}\chi^{1/3}
\end{equation}
Because of the latter our assumption $t\gg m/eE$ corresponds to $\chi\gg 1$.
Proceeding exactly in the same way as in previous case, we now obtain
\begin{equation}\label{WpFrr1}
W_\gamma\simeq \frac{e^2}{t_q}\sim \frac{e^2m^2}{p}\chi^{2/3},\quad F_{RR}\simeq kW_\gamma\sim e^2m^2\chi^{2/3},
\end{equation}
again to be compared with the exact results
\cite{Ritus64,Baier67,Landau4}
\begin{eqnarray}\label{hpe_exact1}
W_\gamma\approx \frac{14\Gamma(2/3)}{27}\frac{e^2m^2}{p}(3\chi)^{2/3}\approx 1.46\frac{e^2m^2}{p}\chi^{2/3},\\\label{hpe_exact2} F_{RR}\approx \frac{32\Gamma(2/3)}{243}e^2m^2(3\chi)^{2/3}\approx 0.37 e^2m^2\chi^{2/3}.
\end{eqnarray}

\section{Pair photoproduction by hard photon}

\begin{figure}[th!]
\includegraphics[width=0.2\textwidth]{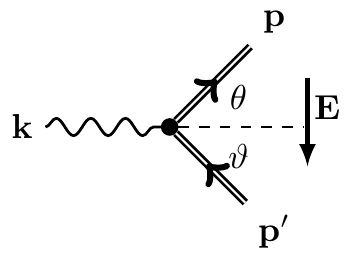}
\caption{\label{Fig4} Feynman diagram for pair photoproduction by a photon in a constant electric field.}
\end{figure}

Pair photoproduction by a photon in the field is often alternatively called non-linear Breit-Wheeler process. Its diagram is represented in Fig.~\ref{Fig4}. Our assumptions and consideration almost repeat the previous section (in particular, we assume $k,p,p'\gg m$), so that let us focus mostly on differences.

Ignoring for the sake of simplicity the transverse momentum of the particles (as its consideration would precisely repeat the photon emission case), and as for the rest proceeding as before, we can start directly with
\begin{eqnarray}
\Delta\varepsilon(t)=\varepsilon_{\vec{k}-\vec{p}}(t)+\varepsilon_{\vec{p}}(t)-k\approx\sqrt{(k-p)^2+e^2E^2t^2+m^2}+\nonumber\\+\sqrt{p^2+e^2E^2t^2+m^2}-k\approx\frac{k\left(e^2E^2t^2+m^2\right)}{2p(k-p)}\gtrsim \frac{2\left(e^2E^2t^2+m^2\right)}{k}\label{de_ppp}
\end{eqnarray}
Again, this is $\Delta\varepsilon=\mathcal{O}(m/\gamma)$, i.e. much smaller than $\mathcal{O}(m)$ for spontaneous pair creation from vacuum, this is exactly the reason why pair photoproduction requires a weaker field to take place. An important but rather obvious kinematical difference from the case of hard photon emission is that now the energy lack is bounded from below (i.e., there exists a threshold), its minimal value is attained when initial momentum is equally distributed among the electron and the positron ($p=p'=k/2$).

Let us proceed directly to limiting cases. In the quantum (or strong field) case $t\gg m/eE$, by neglecting $m^2$ term in the numerator we arrive at
\begin{equation}\label{de_ppp1}
\Delta\varepsilon(t)\approx\frac{e^2E^2t^2}{k}
\end{equation}
The angular spread is estimated by $\theta,\vartheta\simeq eEt/k\ll 1$ and arises again just due to contortion of trajectories in the field. The relation $eE\theta t\simeq\Delta\varepsilon$ is satisfied identically, indicating that strong field always has enough time to make virtual particles real by providing them enough energy.

So that, the probability rate is defined entirely by a `quantum time' $t_q$, which is defined by 
\begin{equation}\label{tq_ppp}
t_q\simeq\frac1{\Delta\varepsilon(t_q)}\simeq\frac{k}{e^2E^2t_q^2}\quad t_q\simeq\left(\frac{k}{e^2E^2}\right)^{1/3}=\frac{k}{m^2}\varkappa^{-2/3}=\frac{m}{eE}\varkappa^{1/3},
\end{equation}
where $\varkappa\gg 1$ is the quantum parameter for the initial photon (analogue of $\chi$). Hence, the probability rate should be of the form
\begin{equation}\label{We}
W_{e^-e^+}\simeq \frac{e^2}{t_q}\sim \frac{e^2m^2}{k}\varkappa^{2/3},
\end{equation}
(the exact result differs only by the numerical factor $15\cdot 3^{2/3} \Gamma^4(2/3)/(28\pi^2)=0.38$ \cite{Ritus64,Baier67,Landau4}).

In the weak field case $t\ll m/eE$ from (\ref{de_ppp}) we have
\begin{equation}\label{de_tq}
\Delta\varepsilon\simeq \frac{m^2}{k},\quad t_q\simeq\frac1{\Delta\varepsilon}\simeq\frac{k}{m^2}.
\end{equation}
Assuming $p=p'=k/2$, the angles can be now estimated by $\theta,\vartheta\simeq m/k\ll 1$, and hence for the time $t_e$ we have
\begin{equation}\label{te_ppp1}
t_e\simeq\frac{\Delta\varepsilon}{eE\vartheta}\simeq\frac{m}{eE}.
\end{equation}
The scalings (\ref{de_tq}), (\ref{te_ppp1}) can be checked for consistency by reconsidering the process in a reference frame $K'$, such that the incoming photon is soft ($k'\simeq m$). Obviously, this RF should move along with the incoming photon with $\gamma$-factor $\gamma\sim k/m$ with respect to the lab frame. Obviously, in $K'$, $\Delta\varepsilon'\simeq m$, $E'\simeq\gamma E$ and $\vartheta'\simeq 1$, hence $t_e'\simeq \Delta\varepsilon'/eE'\simeq m/(eE\gamma)$ and $t_q'\simeq 1/\Delta\varepsilon'\simeq 1/m$. But since $t_{e,q}\simeq \gamma t_{e,q}'$, we arrive at the same results (\ref{de_tq}), (\ref{te_ppp1}) as before.

According to (\ref{de_tq}), (\ref{te_ppp1}) $t_q/t_e\simeq \varkappa$. Since for the case under consideration $\varkappa\ll 1$, we expect that probability of the process would be suppressed by a factor $\mathcal{O}(e^{-t_e/t_q})=\mathcal{O}(e^{-1/\varkappa})$. We can reproduce this factor more precisely by using the prescription (\ref{tr_amp_est}). For this, we first seek for a stationary point of the expression (\ref{de_ppp}),
\begin{equation}\label{sp_ppp}
\Delta\varepsilon(it_*)=\frac{2\left(-e^2E^2t_*^2+m^2\right)}{k}=0\quad\Longrightarrow\quad t_*=\frac{m}{eE},
\end{equation}
(note that again $t_*=t_e$!) and then just evaluate the integral
\begin{eqnarray}
W_{e^-e^+}\propto \left\vert\exp\left(-\int\limits_0^{t_*}\Delta\varepsilon(it)\,dt\right)\right\vert^2=\left\vert\exp\left(-\int\limits_0^{m/eE}\frac{2\left(-e^2E^2t^2+m^2\right)}{k}\,dt\right)\right\vert^2=\nonumber\\
=\left\vert e^{-4m^3/3eEk}\right\vert^2=e^{-8/3\varkappa}.\label{tun_factor}
\end{eqnarray}
Yet again, the result is in full agreement with explicit calculations \cite{Ritus64,Baier67,Landau4}.

\section{Mass operator}

Unfortunately, as for now the simple approach described above can not be readily applied to study all the variety of more complicated processes. The most essential and obvious reasons for this are: (i) absence (or at least as for now not yet developed enough understanding) of classical counterparts for some of the QFT quantities (such as, e.g., polarization operator or the vertex function), and (ii) ambiguity in constructing the probabilities by using exclusively the dimensional arguments (e.g., in case of several time scales at hand, each corresponding e.g. to a particular loop). 

Let us try here just to understand the main features of behavior of the mass operator in external field, which is distinguished by existence of a rather close classical analogue, the self-energy. In the absence of the field, the mass operator in the second order is given by \cite{Landau4}
\begin{equation}\label{mass_op}
M^{(2)}=e^2\int\limits \frac{d^4k}{(2\pi)^4}\gamma_\mu S(p-k)\gamma^\mu D(k)=\vbox{
\begin{tabular}{c}
\includegraphics[width=0.2\columnwidth]{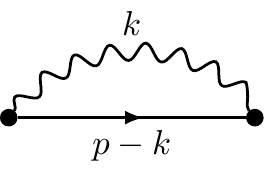}
\end{tabular}},
\end{equation}
and, as is well known, is divergent. To treat it, one ought first to introduce an appropriate regularization, then make computation and mass renormalization, and only afterwords to take the regularization off. Not specifying a particular regularization procedure (which, as is also well known, must preserve as much symmetries of the problem as possible to avoid the false anomalies), assume only that the admissible spatial and momentum resolutions are somehow restricted by $r\gtrsim R$ and $k\lesssim R^{-1}$, where $R$ is the regularization parameter (which can be roughly thought of as the electron radius\footnote{The regularization parameter should be defined invariantly, so that it should in fact better correspond to electron size in a proper reference frame. In order to avoid this complication, we assume for a while that the initial electron is at rest.\label{note}}).

\begin{figure}[th!]
\includegraphics[width=0.7\textwidth]{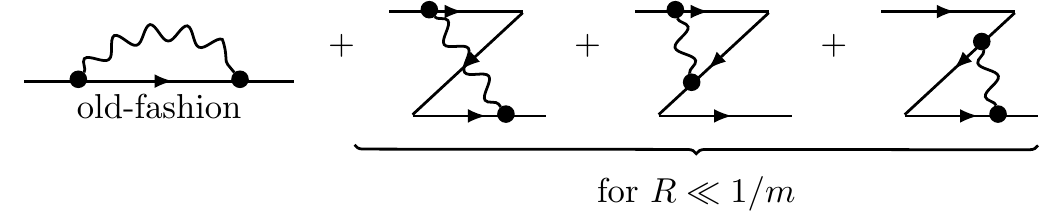}
\caption{\label{Fig5} A set of old-fashion PT diagrams for the mass operator.}
\end{figure}

As is known, in a classical theory the self-energy diverges as $M_C\simeq e^2/R$ as $R\to 0$, while in QED this divergence becomes much weaker, $M_{QED}^{(2)}\simeq e^2m\ln[1/(mR)]$ \cite{Weisskopf,Landau4}. This difference can be well understood in a pictorial way in the framework of the old-fashion perturbation theory (see Fig.~\ref{Fig5}).
As shown in the figure, the usual Feynman propagator corresponds to propagation of both virtual electrons and positrons at the same time (the latter correspond to such paths that are directed backwards in time). In particular, it includes the zigzag-like paths corresponding to creation and further annihilation of virtual pairs on the way. So that, besides just the ordinary propagation with emittance and further absorbtion of a virtual photon (which contributes as $\sim e^2/R$), in case $R\lesssim 1/m$ there also come other comparable contributions with creation and further annihilation of one (shown by the inclined lines) or even more (not shown) virtual positrons. These virtual electrons and positrons also interact with the initial electron as well as with each other. In the absence of the field, all the interactions shown in Fig.~\ref{Fig5} can be estimated as $\sim\pm e^2/R$, where plus/minos sign refers to interaction between the electrons or between an electron and the positron, respectively. In total, among the shown diagrams the first two terms come with $+$ and the last two come with $-$, and the same happens for diagrams with a larger number of virtual pairs on the way. This is exactly the reason for cancellation of the dominant divergencies in QED against the classical electrodynamics (in modern notation, transition from inverse proportionality to logarithmic dependence on $R$ as it decreases passing through the value of Compton length $l_C=1/m$ is nicely and explicitly demonstrated in Ref.~\cite{Vilenkin}).

Now, in a view of the above discussion, it becomes more or less clear, that in the presence of a transverse electric field the virtual pairs should polarize, with positrons and electrons shifting apart (as positrons are
shifted along while electrons opposite to the field), thus violating the mentioned perfect cancellation of dominating divergencies. As a result, it is expected, that at least in case of strong enough field, the inverse power dependence on $R$ should be somehow recovered back.
\begin{figure}[th!]
\includegraphics[width=0.3\textwidth]{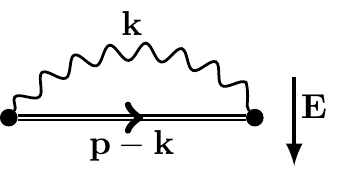}
\caption{\label{Fig6} An IFQED Feynman diagram for the mass operator in a transverse electric field.}
\end{figure}

For the mass operator, the renormalization procedure implies subtraction the field-free value. As already discussed in Sec.~\ref{Sec3}, the time of existence of a virtual photon for $p\gg m$ is given by Eq.~(\ref{tq_hpe2}). If we are interested in a real part of the mass operator, the photon never becomes real and hence we expect the time $t_e$ to be irrelevant. For $t\ll t_q$ we also expect that influence of the field is still non-important, since the particles move as if they were free. This means that what we obtain after the subtraction could be very roughly estimated by substitution of $t_q$ instead of $R$. However, we must certainly rescale beforehand the value of $t_q$ into the proper reference frame of the initial electron (see the footnote on page \pageref{note}), $\tau_q\simeq t_q/\gamma\simeq 1/(m\chi^{2/3})$. Thus we arrive at
\begin{equation}\label{mass}
M\simeq \frac{e^2}{\tau_q}\simeq \alpha m\chi^{2/3},\quad\chi\gg 1.
\end{equation}
As before, this reproduces the exact result \cite{Ritus70} apart only from the numerical coefficient $(28/27)\pi\cdot 3^{1/6}\Gamma(2/3)\approx 5.3$. For the opposite weak field limit $\chi\ll 1$, Ritus provides the leading asymptotic expression ${\rm Re}\,M^{(2)}\simeq\frac{16}{3}e^2m\chi^2\ln{\frac1{\chi}}$, which, contrary to the just considered case $\chi\gg 1$, contains a logarithmic factor. This feature could be expected from the above discussion, however as for now I have not yet understood fully enough how this expression could be revealed completely within the approach presented here. Coming back to the case $\chi\gg 1$, note that the transverse size of the loop can be estimated as 
\begin{equation}\label{r_perp}
r_\perp\sim t_q\vartheta\sim \frac{eEt_q^2}{p}\sim \frac1{(eEp)^{1/3}}\sim \frac1{m}\chi^{-1/3}.
\end{equation}
Interestingly, it is mass-independent, as was the time $\tilde{t}_q\simeq (eE)^{-1/2}$ in case of spontaneous pair creation from vacuum.

The validity of perturbation theory implies smallness of radiation corrections, in particular $M\ll m$. This suggests that the actual expansion parameter for IFQED in case $\chi\gg 1$ is $\alpha\chi^{2/3}$ rather than just $\alpha$. This important conjecture was in fact made already quite long ago and confirmed by direct computation of all the leading diagrams up to 6th order \cite{Narozhny80}. Practically, this conjecture (if true) would imply a qualitative difference between the ordinary QED and the IFQED -- the former remains a perturbative QFT at least as soon as we are away from the Landau pole (which corresponds to unphysically huge energy), while the latter would become non-perturbative already at $\chi\simeq 10^3$. For transverse motion of particles in the laser field this threshold is illustrated for better reference in Table \ref{tab}. In particular, one can observe that this threshold could be almost overcome experimentally by combining the most powerful already existing laser systems with the future International Linear Collider (ILC)--class TeV lepton colliders.

\begin{table}[th!]
\begin{center}
\caption{Typical bench values of the energies of initial particles and the corresponding field strength and laser intensities implying attaining the non-perturbative regime of IFQED ($\alpha\chi^{2/3}\simeq 1$).\label{tab}}
\setlength{\tabcolsep}{12pt}
 \renewcommand{\arraystretch}{1.5}
\begin{tabular}{|l|l|l|l|l|}
\hline
$\varepsilon_{in}$, GeV &  $10^3$&  $10^2$&  $10$&  $1$\\ \hline
$E/E_S$ & $10^{-3}$ & $10^{-2}$ & $0.1$  & $1$ \\ \hline
$I_L$,W/cm$^2$ & $10^{23}$ & $10^{25}$ & $10^{27}$ & $10^{29}$ \\ \hline
\end{tabular}
\end{center}
\end{table}

To avoid possible confusion, note that the data presented in Table \ref{tab} necessarily reefers to the just very case of transverse propagation of particles in the field. For example, for self-sustained cascades, as shown in Ref.~\cite{Fedotov10}, the particles are accelerated entirely by the field itself, thus always keeping small angle with it. Because of that, in that case the parameter $\chi$ can be estimated as $\chi\simeq (E/\alpha E_S)^{3/2}$, so that $\alpha\chi^{2/3}\simeq E/E_S\ll 1$ as long as the case of subcritical field is considered (for recent debates about the very attainability of the critical field $E_S$ in laser-matter interactions we also refer to the same Ref.~\cite{Fedotov10}).

\section{Conclusion}

Because of fast growth of the number of theoretical papers on IFQED, as well as in a view of the increasing number of experimental proposals, it is now becoming of real importance to deep our qualitative understanding of the nature of IFQED processes, in particular such their key parameters as formation time/length scales. This would be especially important for resolution of such old but (in my opinion) still not fully addressed problems as e.g. possible emergence of non-perturbative nature of IFQED \cite{Narozhny80} or interplay between the short distance/high energy behavior of QED and the strong field limit of IFQED \cite{Ritus75}.

My purpose was to attract the attention of the community to this challenging task by demonstrating how at least some known results could be reproduced almost without calculations. I realize, however, that even the examples I have considered here are not yet presented in a final refined enough manner and may rise questions (at least, I am still not enough satisfied with some points). Still, my hope is that these notes will be still useful for further progress.

These notes represent a talk given recently at the Conference on Extremely High Intensity Laser Physics (ExHILP) in Heidelberg. I would like to thank the organizers Antonino Di Piazza, Karen Z. Hatsagortsyan and  Christoph H. Keitel for their kind invitation. This work was supported by the Russian Fund for Basic Research (grant 13-02- 00372) and the RF President program for support of the leading research schools (grant NSh-4829.2014.2).


\begin{thebibliography}{99}
\bibitem{diPiazza12}A. Di Piazza, C. M\"uller, K.Z. Hatsagortsyan, and C.H. Keitel, Rev. Mod. Phys. {\bf 84}, 1177 (2012).
\bibitem{Narozhny15}N.B. Narozhny, and A.M. Fedotov, Contemporary Physics {\bf 56}, 249 (2015).
\bibitem{SLAC1}D. Burke et al., Phys. Rev. Lett. {\bf 79}, 1626  (1997). 
\bibitem{SLAC2}C. Bamber et al., Phys. Rev. D {\bf 60}, 092004 (1999).
\bibitem{Migdal72}A. B. Migdal, Sov. Phys. JETP {\bf 35}, 845 (1972).
\bibitem{Akhmedov11}E. Kh. Akhmedov, Physics of Atomic Nuclei {\bf 74}, 1299 (2011) [arXiv:1011.3776].
\bibitem{Schwartz14}M.D. Schawrtz, {\it Quantum Field Theory and the Standard Model}, (Cambridge University Press, Cambridge, 2014).
\bibitem{Popov}V.S. Popov, Pis'ma v ZhETP {\bf 13}, 261 (1971).
\bibitem{Nikishov79}A.I. Nikishov, Trudy FIAN, Vol. 111, pp. 152-271 (1979).
\bibitem{Ritus64}A. I. Nikishov, V. I. Ritus, Sov. Phys. JETP  {\bf 19}, 529 (1964); {\bf25}, 1135 (1967).
\bibitem{Baier67}V. N. Baier, V. M. Katkov, Sov. Phys. JETP {\bf 26}, 854 (1968).
\bibitem{Landau4}E.~M. Lifshitz, L.~P. Pitaevskii, and V.~B. Berestetskii, {\it Landau-Lifshitz Course of Theoretical Physics, Vol. 4: Quantum Electrodynamics}, (Reed Educational and Professional Publishing, Oxford, 1982).
\bibitem{Weisskopf}V. Weisskopf, Zeitschrift f\"ur Physik {\bf 89}, 27 (1934).
\bibitem{Vilenkin} A.V. Vilenkin and P.I. Fomin, Sov. Phys. JETP {\bf 40}, 6 (1974).
\bibitem{Ritus70}V.I. Ritus, Zh. Eksp. Teor. Fiz. {\bf 30}, 1181 (1970).
\bibitem{Narozhny80}N.B. Narozhny, Phys. Rev. D {\bf 21}, 1176 (1980).
%
\bibitem{Fedotov10}A. M. Fedotov, N. B. Narozhny, G. Mourou, and G. Korn, Phys. Rev. Lett. {\bf 105}, 080402 (2010).
\bibitem{Ritus75}V.I. Ritus, Zh. Eksp. Teor. Fiz. {\bf 69}, 1517 (1975).
\end{thebibliography}
\end{document}